\documentclass[twocolumn,letterpaper,aps,prd,longbibliography,superscriptaddress,floatfix]{revtex4-1}

\usepackage{graphicx}	% Include figure files
\usepackage{xspace}     % Include xspace

\begin{document}

%%%%%%%%%%%%%%%%%%%%%%%%%%%%%%%%%%%%%%%%%%%%%%%%% Title of paper

\title{Transverse momentum dependent forward neutron single spin 
asymmetries in transversely polarized $p$$+$$p$ collisions at 
$\sqrt{s}=200$~GeV }

\newcommand{\abilene}{Abilene Christian University, Abilene, Texas 79699, USA}
\newcommand{\augie}{Department of Physics, Augustana University, Sioux Falls, South Dakota 57197, USA}
\newcommand{\banaras}{Department of Physics, Banaras Hindu University, Varanasi 221005, India}
\newcommand{\barc}{Bhabha Atomic Research Centre, Bombay 400 085, India}
\newcommand{\baruch}{Baruch College, City University of New York, New York, New York, 10010 USA}
\newcommand{\bnlcoll}{Collider-Accelerator Department, Brookhaven National Laboratory, Upton, New York 11973-5000, USA}
\newcommand{\bnlphys}{Physics Department, Brookhaven National Laboratory, Upton, New York 11973-5000, USA}
\newcommand{\caucr}{University of California-Riverside, Riverside, California 92521, USA}
\newcommand{\charlesczech}{Charles University, Ovocn\'{y} trh 5, Praha 1, 116 36, Prague, Czech Republic}
\newcommand{\cns}{Center for Nuclear Study, Graduate School of Science, University of Tokyo, 7-3-1 Hongo, Bunkyo, Tokyo 113-0033, Japan}
\newcommand{\colorado}{University of Colorado, Boulder, Colorado 80309, USA}
\newcommand{\columbia}{Columbia University, New York, New York 10027 and Nevis Laboratories, Irvington, New York 10533, USA}
\newcommand{\czechtech}{Czech Technical University, Zikova 4, 166 36 Prague 6, Czech Republic}
\newcommand{\debrecen}{Debrecen University, H-4010 Debrecen, Egyetem t{\'e}r 1, Hungary}
\newcommand{\elte}{ELTE, E{\"o}tv{\"o}s Lor{\'a}nd University, H-1117 Budapest, P{\'a}zm{\'a}ny P.~s.~1/A, Hungary}
\newcommand{\eszterhazy}{Eszterh\'azy K\'aroly University, K\'aroly R\'obert Campus, H-3200 Gy\"ongy\"os, M\'atrai \'ut 36, Hungary}
\newcommand{\ewha}{Ewha Womans University, Seoul 120-750, Korea}
\newcommand{\famu}{Florida A\&M University, Tallahassee, FL 32307, USA}
\newcommand{\fsu}{Florida State University, Tallahassee, Florida 32306, USA}
\newcommand{\gsu}{Georgia State University, Atlanta, Georgia 30303, USA}
\newcommand{\hiroshima}{Hiroshima University, Kagamiyama, Higashi-Hiroshima 739-8526, Japan}
\newcommand{\howard}{Department of Physics and Astronomy, Howard University, Washington, DC 20059, USA}
\newcommand{\ihepprot}{IHEP Protvino, State Research Center of Russian Federation, Institute for High Energy Physics, Protvino, 142281, Russia}
\newcommand{\illuiuc}{University of Illinois at Urbana-Champaign, Urbana, Illinois 61801, USA}
\newcommand{\inrras}{Institute for Nuclear Research of the Russian Academy of Sciences, prospekt 60-letiya Oktyabrya 7a, Moscow 117312, Russia}
\newcommand{\instpasczech}{Institute of Physics, Academy of Sciences of the Czech Republic, Na Slovance 2, 182 21 Prague 8, Czech Republic}
\newcommand{\isu}{Iowa State University, Ames, Iowa 50011, USA}
\newcommand{\jaea}{Advanced Science Research Center, Japan Atomic Energy Agency, 2-4 Shirakata Shirane, Tokai-mura, Naka-gun, Ibaraki-ken 319-1195, Japan}
\newcommand{\jeonbuk}{Jeonbuk National University, Jeonju, 54896, Korea}
\newcommand{\kek}{KEK, High Energy Accelerator Research Organization, Tsukuba, Ibaraki 305-0801, Japan}
\newcommand{\korea}{Korea University, Seoul 02841, Korea}
\newcommand{\kurchatov}{National Research Center ``Kurchatov Institute", Moscow, 123098 Russia}
\newcommand{\kyoto}{Kyoto University, Kyoto 606-8502, Japan}
\newcommand{\lawllnl}{Lawrence Livermore National Laboratory, Livermore, California 94550, USA}
\newcommand{\losalamos}{Los Alamos National Laboratory, Los Alamos, New Mexico 87545, USA}
\newcommand{\lund}{Department of Physics, Lund University, Box 118, SE-221 00 Lund, Sweden}
\newcommand{\lyon}{IPNL, CNRS/IN2P3, Univ Lyon, Université Lyon 1, F-69622, Villeurbanne, France}
\newcommand{\maryland}{University of Maryland, College Park, Maryland 20742, USA}
\newcommand{\mass}{Department of Physics, University of Massachusetts, Amherst, Massachusetts 01003-9337, USA}
\newcommand{\michigan}{Department of Physics, University of Michigan, Ann Arbor, Michigan 48109-1040, USA}
\newcommand{\muhlenberg}{Muhlenberg College, Allentown, Pennsylvania 18104-5586, USA}
\newcommand{\nara}{Nara Women's University, Kita-uoya Nishi-machi Nara 630-8506, Japan}
\newcommand{\natmephi}{National Research Nuclear University, MEPhI, Moscow Engineering Physics Institute, Moscow, 115409, Russia}
\newcommand{\newmex}{University of New Mexico, Albuquerque, New Mexico 87131, USA}
\newcommand{\nmsu}{New Mexico State University, Las Cruces, New Mexico 88003, USA}
\newcommand{\northcg}{Physics and Astronomy Department, University of North Carolina at Greensboro, Greensboro, North Carolina 27412, USA}
\newcommand{\ohio}{Department of Physics and Astronomy, Ohio University, Athens, Ohio 45701, USA}
\newcommand{\ornl}{Oak Ridge National Laboratory, Oak Ridge, Tennessee 37831, USA}
\newcommand{\orsay}{IPN-Orsay, Univ.~Paris-Sud, CNRS/IN2P3, Universit\'e Paris-Saclay, BP1, F-91406, Orsay, France}
\newcommand{\peking}{Peking University, Beijing 100871, People's Republic of China}
\newcommand{\pnpi}{PNPI, Petersburg Nuclear Physics Institute, Gatchina, Leningrad region, 188300, Russia}
\newcommand{\pusan}{Pusan National University, Pusan 46241, Korea}
\newcommand{\riken}{RIKEN Nishina Center for Accelerator-Based Science, Wako, Saitama 351-0198, Japan}
\newcommand{\rikjrbrc}{RIKEN BNL Research Center, Brookhaven National Laboratory, Upton, New York 11973-5000, USA}
\newcommand{\rikkyo}{Physics Department, Rikkyo University, 3-34-1 Nishi-Ikebukuro, Toshima, Tokyo 171-8501, Japan}
\newcommand{\saispbstu}{Saint Petersburg State Polytechnic University, St.~Petersburg, 195251 Russia}
\newcommand{\seoulnat}{Department of Physics and Astronomy, Seoul National University, Seoul 151-742, Korea}
\newcommand{\stonybrkc}{Chemistry Department, Stony Brook University, SUNY, Stony Brook, New York 11794-3400, USA}
\newcommand{\stonycrkp}{Department of Physics and Astronomy, Stony Brook University, SUNY, Stony Brook, New York 11794-3800, USA}
\newcommand{\tenn}{University of Tennessee, Knoxville, Tennessee 37996, USA}
\newcommand{\titech}{Department of Physics, Tokyo Institute of Technology, Oh-okayama, Meguro, Tokyo 152-8551, Japan}
\newcommand{\tsukuba}{Tomonaga Center for the History of the Universe, University of Tsukuba, Tsukuba, Ibaraki 305, Japan}
\newcommand{\vandy}{Vanderbilt University, Nashville, Tennessee 37235, USA}
\newcommand{\weizmann}{Weizmann Institute, Rehovot 76100, Israel}
\newcommand{\wigner}{Institute for Particle and Nuclear Physics, Wigner Research Centre for Physics, Hungarian Academy of Sciences (Wigner RCP, RMKI) H-1525 Budapest 114, POBox 49, Budapest, Hungary}
\newcommand{\yonsei}{Yonsei University, IPAP, Seoul 120-749, Korea}
\newcommand{\zagreb}{Department of Physics, Faculty of Science, University of Zagreb, Bijeni\v{c}ka c.~32 HR-10002 Zagreb, Croatia}
\affiliation{\abilene}
\affiliation{\augie}
\affiliation{\banaras}
\affiliation{\barc}
\affiliation{\baruch}
\affiliation{\bnlcoll}
\affiliation{\bnlphys}
\affiliation{\caucr}
\affiliation{\charlesczech}
\affiliation{\cns}
\affiliation{\colorado}
\affiliation{\columbia}
\affiliation{\czechtech}
\affiliation{\debrecen}
\affiliation{\elte}
\affiliation{\eszterhazy}
\affiliation{\ewha}
\affiliation{\famu}
\affiliation{\fsu}
\affiliation{\gsu}
\affiliation{\hiroshima}
\affiliation{\howard}
\affiliation{\ihepprot}
\affiliation{\illuiuc}
\affiliation{\inrras}
\affiliation{\instpasczech}
\affiliation{\isu}
\affiliation{\jaea}
\affiliation{\jeonbuk}
\affiliation{\kek}
\affiliation{\korea}
\affiliation{\kurchatov}
\affiliation{\kyoto}
\affiliation{\lawllnl}
\affiliation{\losalamos}
\affiliation{\lund}
\affiliation{\lyon}
\affiliation{\maryland}
\affiliation{\mass}
\affiliation{\michigan}
\affiliation{\muhlenberg}
\affiliation{\nara}
\affiliation{\natmephi}
\affiliation{\newmex}
\affiliation{\nmsu}
\affiliation{\northcg}
\affiliation{\ohio}
\affiliation{\ornl}
\affiliation{\orsay}
\affiliation{\peking}
\affiliation{\pnpi}
\affiliation{\pusan}
\affiliation{\riken}
\affiliation{\rikjrbrc}
\affiliation{\rikkyo}
\affiliation{\saispbstu}
\affiliation{\seoulnat}
\affiliation{\stonybrkc}
\affiliation{\stonycrkp}
\affiliation{\tenn}
\affiliation{\titech}
\affiliation{\tsukuba}
\affiliation{\vandy}
\affiliation{\weizmann}
\affiliation{\wigner}
\affiliation{\yonsei}
\affiliation{\zagreb}
\author{U.A.~Acharya} \affiliation{\gsu} 
\author{C.~Aidala} \affiliation{\michigan} 
\author{Y.~Akiba} \email[PHENIX Spokesperson: ]{akiba@rcf.rhic.bnl.gov} \affiliation{\riken} \affiliation{\rikjrbrc} 
\author{M.~Alfred} \affiliation{\howard} 
\author{V.~Andrieux} \affiliation{\michigan} 
\author{N.~Apadula} \affiliation{\isu} 
\author{H.~Asano} \affiliation{\kyoto} \affiliation{\riken} 
\author{B.~Azmoun} \affiliation{\bnlphys} 
\author{V.~Babintsev} \affiliation{\ihepprot} 
\author{N.S.~Bandara} \affiliation{\mass} 
% ...
\collaboration{PHENIX Collaboration} \noaffiliation

\date{\today}

%------------------------------------------------------------------------------|

\begin{abstract}

In 2015, the PHENIX collaboration has measured very forward ($\eta > 
6.8$) single spin asymmetries of inclusive neutrons in transversely 
polarized proton-proton and proton-nucleus collisions at a center of 
mass energy of 200 GeV. A previous publication from this data set 
concentrated on the nuclear dependence of such asymmetries. In this 
measurement the explicit transverse momentum dependence of inclusive 
neutron single spin asymmetries for proton-proton collisions is 
extracted using a bootstrapping unfolding technique on the transverse 
momenta. This explicit transverse momentum dependence will help improve 
the understanding of the mechanisms that create these asymmetries.  

\end{abstract}

\maketitle

%%%%%%%%%%%%%%%%%%%%%%%%%%%%%%%%%%%%%%%%%%%%%%%%%%%%%%%%%%% Introduction
\section{Introduction}

At the beginning of the era of polarized proton collisions at the 
relativistic heavy ion collider (RHIC), a dedicated experiment based on 
a prototype zero-degree calorimeter (ZDC)~\cite{Adler:2000bd} was set up 
to initially study very forward neutral pion asymmetries in transversely 
polarized proton collisions in relation to earlier results that showed 
nonzero results~\cite{Adams:1991rw}.  Instead of finding a neutral pion 
asymmetry, that was only recently discovered at low transverse momentum 
by the RHICf experiment~\cite{Kim:2020sxu}, a sizable neutron asymmetry 
was found in the forward direction of the transversely polarized proton 
beam~\cite{Bazilevsky:2006vd}.

Earlier theoretical studies related very forward neutron production to 
the one-pion-exchange (OPE) 
model~\cite{Soffer:1991am,DAlesio:1998uav,Kopeliovich:1996iw} in which 
the exchange of one pion between the proton and another colliding 
particle can create the outgoing neutron. Such a model was reasonably 
successful in describing unpolarized, very forward neutron production as 
previously observed at the ISR~\cite{Flauger:1976ju}. However, a simple 
pion exchange model would not be able to describe any spin dependence of 
the observed neutron distributions. To accommodate that, an interference 
with another particle exchange would be necessary to have 
helicity-flip and nonflip amplitudes available that can create a single 
spin left-right asymmetry. Within the general framework of Regge 
theory~\cite{Collins:1977jy}, such an interference could be accomplished 
when adding also a scalar meson exchange and the resulting 
pseudoscalar-scalar meson interference would then create the asymmetry. 
Recent calculations of such an OPE based description of very forward 
neutron single spin asymmetries~\cite{Kopeliovich:2011bx} are able to 
qualitatively describe the RHIC measurements which, by now, include 
transversely polarized proton-proton collisions at $\sqrt{s}= $ 62 GeV, 
200 GeV, as well as 500 GeV~\cite{Adare:2012vw}, although the transverse 
momentum information enters only indirectly via the different collision 
energies.

The very different asymmetries observed in proton-nucleus 
collisions~\cite{Aidala:2017cnz}, with different sign and much larger 
magnitude, indicate that at high impact parameters and at least for 
high-Z nuclei ultraperipheral collisions (UPC)~\cite{Mitsuka:2017czj} 
also contribute to these asymmetries in a very different way. This data 
together with the recent very forward nonzero neutral pion 
result~\cite{Kim:2020sxu} may provide crucial information to the 
underlying mechanisms that create these asymmetries.

So far, none of these results have been extracted with an explicit 
transverse momentum dependence while the different collision energies 
provide some indirect information on it. Obtaining it can directly test 
the proposed mechanism and the dependence that results from its theory 
calculation~\cite{Kopeliovich:2011bx}. Therefore, extracting the actual 
transverse momentum dependence is the focus of this publication. A 
substantial understanding is required of the transverse momentum 
smearing in the PHENIX ZDCs.  Also needed are determinations of 
systematic uncertainties in unfolding transverse momenta, which were 
studied via a Monte Carlo (MC) bootstrap method as described later. 

In the following sections, the detector description, analyzed data sets 
and the forward neutron selection are covered. Next are described the 
procedure for unfolding the neutron single spin asymmetries as a 
function of the reconstructed transverse momenta to obtain the true 
transverse momentum dependence.  Then, the final results are presented 
before summarizing.

%%%%%%%%%%%%%%%%%%%%%%%%%%%%%%%%%%%%%%%%%%%%%%%%%%% Data sets
\section{Data sets}

In 2015, the PHENIX experiment recorded polarized proton-proton and 
proton-nucleus collision data at a center of mass energy of 200 GeV. In 
the proton-proton collision data, the beams were transversely polarized 
with the spin direction pointing vertically up or down with respect to 
the plane defined by the accelerator ring. 

Inclusive neutrons were detected with the ZDC, which comprise 3 modules 
of Cu-W alloy absorbers layered with optical fibers of 1.7 nuclear 
interaction lengths each (51 radiation lengths per module), covering a 
projected area of 10 cm by 10 cm transverse to the beam direction. The 
absorber layers of the ZDC are tilted 45 degrees upward to 
maximize the light yield from \v{C}erenkov light. The location of the 
ZDCs is 18 m up and downstream of the PHENIX beam interaction point, 
thus covering a range of pseudorapidity $\eta > 6.8$. The ZDC is used to 
measure the energy of forward neutrons, and its energy resolution is 
about 20\% for neutron energies of 100 GeV.  Between the first and the 
second module, approximately at the position of the maximal hadronic 
shower are located scintillator strip detectors with a projected width 
of 15 mm horizontally and vertically.  This shower max detector (SMD) is 
used to determine the position of the neutrons that are selected by 
calculating the weighted average of the deposited energy for all strip 
positions. The position resolution of the SMD for neutrons is 
$\approx1$~cm. Additionally, the SMD is used for local polarimetry of 
the polarized beams by making use of the nonzero neutron asymmetries in 
proton-proton collisions and allowed to track the transverse spin 
orientation or confirm the spin orientation to be rotated in the 
longitudinal direction. See Ref.~\cite{Adcox:2003zm} for a
more detailed detector description.

Collision events were selected for this result by a logical OR of north 
and south ZDCs that require approximately an energy deposit of more than 
15 GeV on either detector.  Within the proton collisions were 
accumulated about 35M neutron events that were triggered by the ZDCs.

%%%%%%%%%%%%%%%%%%%%%%%%%%%%%%%% Event and particle selection criteria
\section{Event and particle selection criteria}

Neutron candidates in the north ZDC were selected by requiring more than 
3\% of the total deposited energy to be in the second ZDC module. This 
effectively rejects photon candidates, that deposited their energy in 
the first module due to being electromagnetic showers.  Also nonzero 
hits in both horizontal and vertical SMDs are required to 
reliably estimate the neutron position and transverse momentum. 
Furthermore, the reconstructed neutron energies were selected between 40 
and 120 GeV. The hit position as defined by the SMDs has to be within 
0.5 to 4 cm in radius from the nominal beam position. Additionally, data 
under stable running conditions with no problems in the polarized beam 
diagnostics were selected. The transverse momentum $P_T$ is 
reconstructed from the neutron energy $E$, the radius of the average hit 
position $r$ and the distance from the interaction point $z_{ZDC}$:
\begin{equation}
    P_{T} = \frac{r}{z_{ZDC} }E.
\end{equation}

\noindent The events that fulfill the above conditions are then binned 
in 4 transverse momentum bins of [0.01, 0.06, 0.11, 0.16, 0.21] GeV/$c$ 
and 6 equidistant azimuthal angular bins that cover full azimuth around 
the polarized beam direction. The two spin states are kept separated for 
the unfolding, but to obtain the asymmetries needed in the 
bootstrap method of the MC, they are also directly calculated here as:
\begin{equation}
A_N (\phi) = \frac{1}{\langle P \rangle}\frac{N^+(\phi) - \mathcal{R} N^-(\phi)}{N^+(\phi) +\mathcal{R}N^-(\phi)},
\end{equation}

\noindent where $\langle P\rangle$ is the average beam polarization 
(for this running period 52\%~\cite{polarimetry}) and $N^{\pm}$ are the 
yields of neutrons in the up/down spin state as a function of azimuthal 
angle $\phi$ that is defined relative to the spin-up direction. 
$\mathcal{R}$ is the ratio of accumulated luminosities for the down and 
up spin states, and is close to unity in this analysis. The actual $A_N$ 
is then calculated by fitting a sine modulation to it with magnitude and 
phase as free parameters.

As systematic uncertainties, the amount of charged particle background 
(dominated by protons) and the uncertainty of the beam center position 
need to be evaluated. Unlike other years, no charge veto counter in 
front of the ZDC was installed in this running period. This resulted in 
a rather asymmetric charged hadron background predominantly from protons 
that are swept into the ZDC by the dipole magnet which joins and 
separates the two beams. The fraction of charged hadron background was 
statistically subtracted on the spin dependent yield level by applying 
the background fractions that were obtained in the 2008 running period 
when the charge veto counter was installed. The statistical 
uncertainties on these background fractions were then assigned as 
systematic uncertainties on the resulting raw asymmetries. 

The central beam position relative to the ZDC also cannot be perfectly 
determined due to the large lever arm as well as varying beam 
conditions. As such, the assigned beam position was artificially varied 
by 1 cm horizontally and 0.5 cm vertically around the nominal beam 
position, respectively. All neutron positions, transverse momenta, 
and azimuthal angles were recalculated before evaluating the 
asymmetries. These variations were motivated by the uncertainties based 
on two independent methods of reconstructing the beam center positions 
using the ZDCs.  The combined uncertainties on the asymmetries from 
charged background and beam position are then used as a basis for 
variation of the bootstrap MC method of unfolding the asymmetries as
discussed in the next section.  The uncertainties due to charged 
background and beam position remain negligible compared to the large 
systematic uncertainties this unfolding introduces.

%%%%%%%%%%%%%%%%%%%%%%%%%%%%%%%%%%%%%% Transverse momentum unfolding
\section{Transverse momentum unfolding}

As hadronic showers develop a substantial size and the segmentation of 
the SMDs is limited, the reconstructed neutron energy and in particular 
the transverse momentum are smeared. The transverse momentum dependence 
of the single spin asymmetries is however of much interest for the 
understanding of the mechanism that creates these asymmetries.

We have performed detailed MC simulations using 5 different types of 
event generators as input to full {\sc geant3}~\cite{Brun:1994aa} 
simulations of the forward region of 
PHENIX~\cite{Togawa:2008cca,Adare:2013ekj} including the ZDCs, SMDs, the 
beam-pipe as well as the dipole magnet that merges and separates the 
incoming and outgoing beams and is responsible for an asymmetric spray 
from charged particles. These {\sc geant} simulations have been shown to 
describe these effects, as well as differences between top and bottom 
that originate from the light collection and back scattering in the top 
part of the ZDC. The composition, energy and momentum distributions 
of particles in the far forward region are not very well understood 
in general, and therefore different types of generators were used to 
gauge the impact of these differences on the unfolded asymmetries. 
The three full generators 
{\sc pythia6.1}~\cite{Sjostrand:2001yu}, 
{\sc pythia8.2}~\cite{Sjostrand:2014zea} and 
{\sc dpmjet3.1}~\cite{Roesler:2000he} were applied, where in particular 
diffractive processes are handled very differently.

Additionally, an empirical distribution of forward neutrons in 
longitudinal and transverse momentum was used to mimic an OPE model. In 
this case, a pion that balances the energy and momentum between the 
incoming proton and the thrown neutron (i.e. $p \rightarrow \pi^+ + n$) 
was collided with the other beam using {\sc pythia8} again (i.e. 
$\pi+p$). Moreover, as the forward pA results have indicated 
\cite{Aidala:2017cnz}, ultraperipheral collisions can also play a role 
in forward neutron production although that will be more prominent in 
proton-nucleus collisions. Therefore, the yield and distribution of 
photons from the other beam was simulated using 
{\sc starlight}~\cite{Klein:2016yzr} and collided with the proton beam 
using {\sc pythia8} again.

As none of these generators is intrinsically spin dependent, spin 
effects ($w$ in the following expression) were simulated by 
reweighting generated events as a function of true transverse momentum 
and azimuthal angle where the spin state was randomly assigned. Three 
different functional forms were used in the reweighting to 
provide as much flexibility as possible for the true transverse momentum 
dependence of the single spin asymmetries. The most general 
parameterization is a 3rd order polynomial in the transverse momentum 
with a vanishing constant term due to the requirement for the asymmetry 
to vanish at zero transverse momentum:
\begin{equation}
w = \left (a \cdot P_{T,g} + b \cdot P_{T,g}^2 
+ c \cdot P_{T,g}^3\right) \sin (\phi_g + \lambda \cdot\pi),  
\end{equation}

\noindent where $P_{T,g}$ and $\phi_g$ are the true transverse momenta 
and azimuthal angles, respectively and $\lambda$ ($\pm1$) is the spin 
state while $a$, $b$ and $c$ are free parameters that are varied. A 
second functional form (with $a$ and $b>0$ free parameters) is based on 
a power-law behavior:
\begin{equation}
w = \left (a \cdot P_{T,g}^b\right) \sin (\phi_g + \lambda \cdot\pi),  
\end{equation}

\noindent and the last parameterization (with $a$ and $b$ free 
parameters) follows an exponential form that eventually reaches an 
asymptotic constant:
\begin{equation}
w =  a \left( 1- {\rm e}^{ b\cdot P_{T,g}}\right) 
\sin (\phi_g + \lambda \cdot\pi).  
\end{equation}
In the power law parameterization, only positive powers are allowed to 
avoid unphysical nonzero asymmetries at zero transverse momentum.

\noindent For each set of parameters, functional form, and MC generator, 
the single spin asymmetries were extracted from the reconstructed 
kinematic variables that included these weights based on the true 
variables.

In a first step, the reconstructed asymmetries that were obtained from 
the data are compared to the reconstructed asymmetries from MC for a 
large number of variations of the parameters. The quality of a set of 
parameters, functional form and MC generator in reproducing the data 
asymmetries was evaluated by calculating the $\chi^2$ between the actual 
data points and the smeared asymmetry points. While the MC statistics 
are generally large enough, many functional forms can describe the data 
within the experimentally measured uncertainties.

%%%%%%%%%%%%%%%%%%%%%%%%%%%%%%%%%%%%%%%%%%%%%%%%%%%%%%%% Fig_1
\begin{figure*}[htb]
\begin{minipage}{0.99\linewidth}
\includegraphics[width=0.325\linewidth]{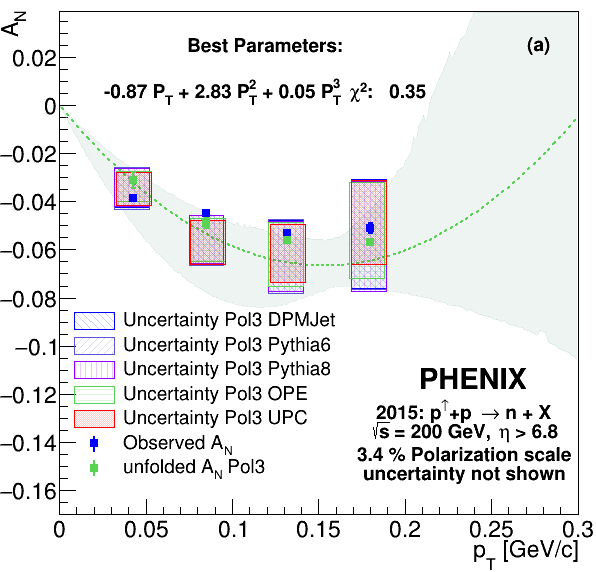}
\includegraphics[width=0.325\linewidth]{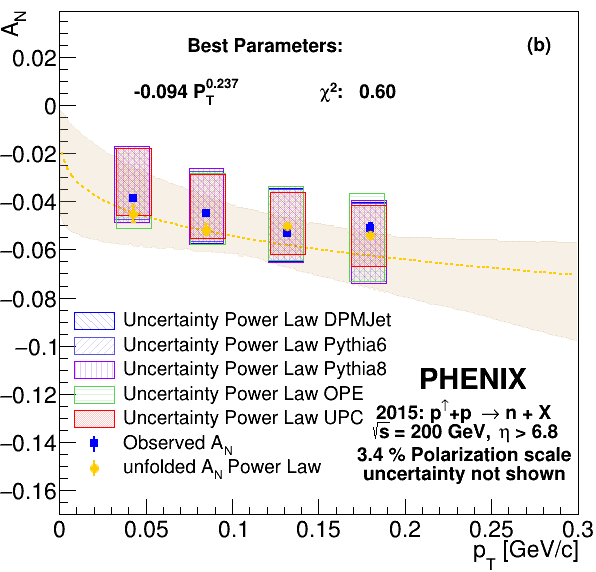}
\includegraphics[width=0.325\linewidth]{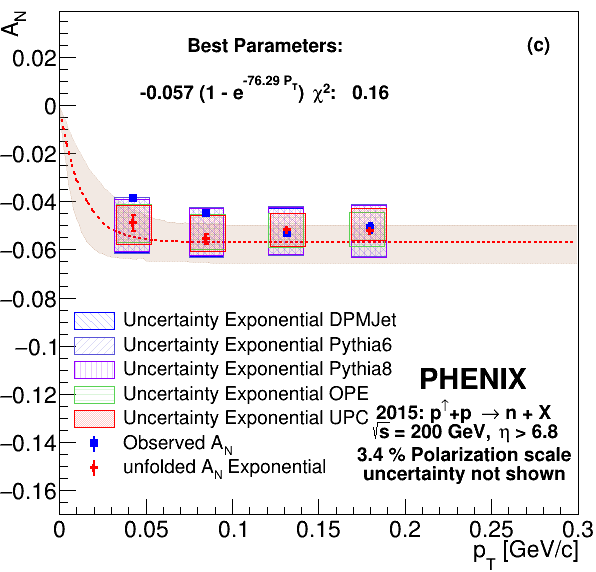}
    \caption{True asymmetry parameterizations as a function of 
transverse momentum for (a) a 3rd order polynomial dependence "Pol3," 
(b) a power-law dependence "Power Law," and (c) an exponential 
dependence "Exponential". The shaded regions represent the regions 
where the $\chi^2$ between the smeared asymmetries related to this 
parameterization and the asymmetries reconstructed from data (solid 
[blue] squares) is below 10 units. The dashed lines represent the 
best matching parameterizations.  Also displayed are the unfolded 
asymmetries (a) solid [dark green] squares, (b) solid [orange] 
circles, and (c) solid [red] hyphens, as obtained from the best 
parameterizations of the OPE generator. The rms ranges of unfolded 
asymmetries are visualized as shaded boxes for the various MC 
generators.}
    \label{fig:parmdist}
\end{minipage}
%\end{figure*}
%%%%%%%%%%%%%%%%%%%%%%%%%%%%%%%%%%%%%%%%%%%%%%%%%%%%% Fig_2
%\begin{figure*}[th]
\begin{minipage}{0.48\linewidth}
\vspace{1.0cm}
    \includegraphics[width=0.99\linewidth]{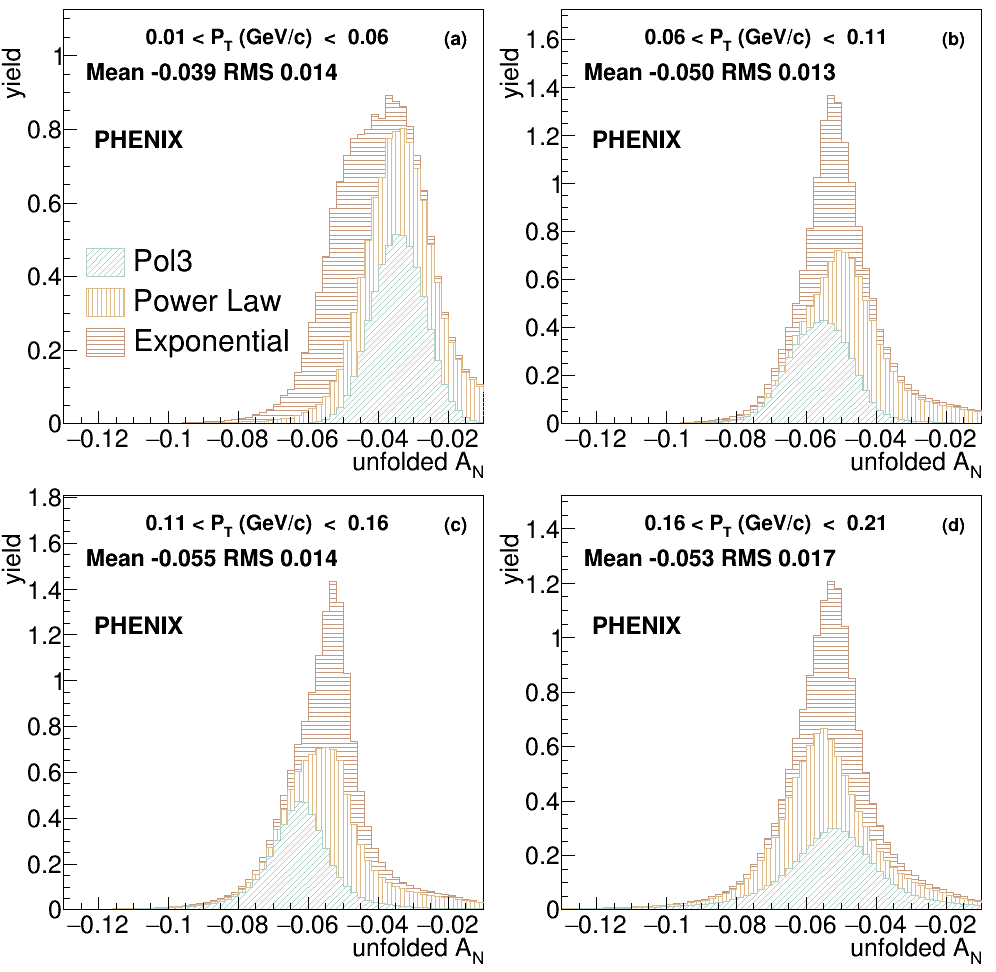}
\caption{Relative likelihood distributions of unfolded asymmetries for 
each transverse momentum bin for all sets of parameters of each 
functional form weighted by the inverse of its $\chi^2$. All different 
MC generators distributions have been combined in these panels. The 
distributions of the 3rd order polynomial parameterization (shaded 
[light green] area "Pol3"), power law behavior (vertically hatched 
[light orange] area "Power Law") and exponential (horizontally hatched 
[light red] area "Exponential") have been stacked in these figures. 
The overall central and rms values are also displayed.
}
    \label{fig:spread}
\end{minipage}
%\end{figure}
%%%%%%%%%%%%%%%%%%%%%%%%%%%%%%%%%%%%%%%%%%%%%%%%%%%%% Fig_3
%\begin{figure}[th]
\hspace{0.2cm}
\begin{minipage}{0.48\linewidth}
\vspace{-0.05cm}
    \includegraphics[width=0.99\linewidth]{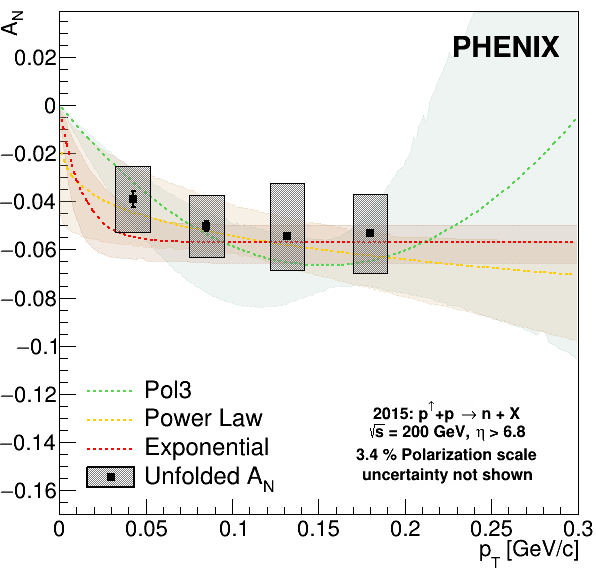}
\vspace{-0.2cm}
\caption{Neutron transverse single spin asymmetries as a function of 
the true transverse momentum. The data points represent the unfolded 
asymmetries obtained via the average over all parameterizations and MC 
generators. The uncertainty boxes represent the systematic uncertainties 
due to the parameterization, functional form, MC generator and unfolding 
procedure.}
    \label{fig:final}
\end{minipage}
\end{figure*}

Figure~\ref{fig:parmdist} displays the functional forms probed for the 
OPE motivated MC generator. The regions where a parametrization results 
in a $\chi^2$ below 10 units is also visualized to highlight the range 
of reasonable parameterizations. Despite the different transverse 
momentum distributions of forward neutrons in the different MC 
generators, their best asymmetry distributions are very similar for each 
set of functional forms. In all functional forms, a rapidly rising 
asymmetry is preferred at small $P_T$, while at intermediate transverse 
momentum (within the second and third data points) no large variation in 
the asymmetries is visible. The exponential function generally prefers 
the asymptotic value to be reached already at these transverse momenta. 
At higher transverse momenta above 0.2 GeV/$c$, the sensitivity is very 
limited for all functional forms despite a nonzero smearing into the 
observed range. As such, the slightly rising, constant or even 
diminishing asymmetries at high transverse momenta can describe the 
measured data reasonably well.

In a second step, the spin-dependent data yields that are 
two-dimensional in reconstructed transverse momentum and azimuthal angle 
are unfolded using the spin-dependent, weighted smearing matrices 
obtained for each set of parameters, functional form and MC generator 
set. For the unfolding itself, the TSVDUnfold package of {\sc root} 
\cite{Brun:1997pa} based on a regularized singular value decomposition 
\cite{Hocker:1995kb} was used. After the unfolding, the asymmetries are 
calculated from the unfolded yields as described above. The best 
parameterization for each functional form and MC generator is used to 
obtain the central point of the unfolded asymmetries and statistical 
uncertainties. The impact of the variation of parameters is evaluated by 
obtaining the root mean square (RMS) of the spread of unfolded 
asymmetries which are weighted by the inverse of their respective 
$\chi^2$ to take the quality of a parameter set into account. 
These uncertainties are also displayed in Fig.~\ref{fig:parmdist}, 
together with the unfolded asymmetries using the OPE generator.

The variation of the unfolded asymmetries is displayed in 
Fig.~\ref{fig:spread} for each transverse momentum bin and 
parameterization, while spreads from the different MC generators had 
been combined equally. The central values from these distributions have 
been taken as the final asymmetry values while the rms value is taken as 
the uncertainty due to the various parameterizations.

In addition to these uncertainties, further systematic uncertainties are 
studied by varying the regularization parameter in the TSVDUnfold method 
as well as the uncertainties due to the statistical uncertainties on the 
smearing matrices themselves. However, most of these values are within 
the boundaries of the uncertainties obtained from the variation of 
parameters and functional forms. Only those systematic contributions 
that exceed the aforementioned uncertainties have been added in 
quadrature.

%%%%%%%%%%%%%%%%%%%%%%%%%%%%%%%%%%%%%%%%%%%%%%% Results
\section{Results}

The inclusive neutron asymmetries obtained from the average of all 
parameterizations and MC generators are displayed in 
Fig.~\ref{fig:final} as a function of the true transverse momentum. 
The final results are tabulated in Table~\ref{tab:final}.

%%%%%%%%%%%%%%5

The absolute values of the asymmetries are consistent with an increase 
with transverse momentum but show an indication of leveling off at 
higher transverse momenta. A simple linear dependence as suggested 
by~\cite{Kopeliovich:2011bx}, as well as the central values of 
transverse momentum integrated asymmetries at different collision 
energies~\cite{Adare:2013ekj,Bazilevsky:2006vd}, seems not necessarily 
to be preferred by the data.  However, a simple linear dependence cannot 
be excluded within uncertainties either. From the MC reweighting 
exercise no substantial differences between the different MC generators 
have been seen.

Taking into account the indication of very different asymmetries in 
ultraperipheral collisions in proton-nucleus 
collisions~\cite{Aidala:2017cnz,Mitsuka:2017czj} and in particular a 
different sign, it appears that the UPC contribution to the 
proton-proton collisions is limited in this $p_T$ region. This is 
expected given the electromagnetic nature of the interaction being 
proportional with $Z^2$. However, in these inclusive results some 
contribution from UPC events may remain, which could alter the 
transverse momentum behavior in comparison to the purely hadronic 
theoretical calculations.

%=================================================== Tble I
\begin{table}[tb!]
\caption{Neutron single spin asymmetries as a function of transverse 
momentum after unfolding transverse-momentum and azimuthal-angular 
smearing. $\Delta A_N$ corresponds to the statistical uncertainties 
while the last two columns specify the upper and lower systematic 
uncertainties $\delta A_N$.}
    \label{tab:final}
\begin{ruledtabular}    \begin{tabular}{c  c  c  r  c}
$\langle P_T \rangle$ (GeV/$c$) &   $A_N$   &   \hphantom{sy}$\Delta A_N$   
& \multicolumn{2}{c}{$\delta A_N$}\\ 
\hline
 0.043 &  -0.039 & $\pm$0.003 & +0.014 & -0.014 \\
 0.085 &  -0.050 & $\pm$0.002 & +0.013 & -0.013 \\
 0.132 &  -0.055 & $\pm$0.002 & +0.022 & -0.014 \\
 0.180 &  -0.053 & $\pm$0.001 & +0.017 & -0.017 \\
    \end{tabular}  \end{ruledtabular}
\end{table}

%%%%%%%%%%%%%%%%%%%%%%%%%%%%%%%%%%%%%%%%%%%%%%%%%% Summary
\section{Summary}

In summary, the PHENIX experiment has measured for the 
first time the transverse momentum dependence of very 
forward neutron single spin asymmetries in proton-proton 
collisions at a center of mass energy of 200 GeV. With 
these measurements the first reliable tests of the 
suggested mechanisms producing such forward neutron 
asymmetries can be performed. While the uncertainties from 
the unfolding are very sizable, a simple linear transverse 
momentum dependence as suggested 
in~\cite{Kopeliovich:2011bx} is not inconsistent; however, 
the asymmetries appear to level off at higher transverse 
momenta. Instead, a much slower rise of the asymmetries or 
even a turnaround at larger transverse momenta is favored 
when considering the best parameterizations.  To understand 
the mechanisms in even more detail, the correlations with 
other detector activity will be useful.

%%%%%%%%%%%%%%%%%%%%%%%%%%%%%%%%%%%%%%%%%%%  Acknowledgements 

%%%%%%%%%%%%%%%%%%%%%%  ACKNOWLEDGMENTS}  %%%%% MGS19 version
%% 2018 change in Korea

\begin{acknowledgments}

We thank the staff of the Collider-Accelerator and Physics
Departments at Brookhaven National Laboratory and the staff of
the other PHENIX participating institutions for their vital
contributions.  We acknowledge support from the
Office of Nuclear Physics in the
Office of Science of the Department of Energy,
the National Science Foundation,
Abilene Christian University Research Council,
Research Foundation of SUNY, and
Dean of the College of Arts and Sciences, Vanderbilt University
(U.S.A),
Ministry of Education, Culture, Sports, Science, and Technology
and the Japan Society for the Promotion of Science (Japan),
Conselho Nacional de Desenvolvimento Cient\'{\i}fico e
Tecnol{\'o}gico and Funda\c c{\~a}o de Amparo {\`a} Pesquisa do
Estado de S{\~a}o Paulo (Brazil),
Natural Science Foundation of China (People's Republic of China),
Croatian Science Foundation and
Ministry of Science and Education (Croatia),
Ministry of Education, Youth and Sports (Czech Republic),
Centre National de la Recherche Scientifique, Commissariat
{\`a} l'{\'E}nergie Atomique, and Institut National de Physique
Nucl{\'e}aire et de Physique des Particules (France),
Bundesministerium f\"ur Bildung und Forschung, Deutscher Akademischer
Austausch Dienst, and Alexander von Humboldt Stiftung (Germany),
J. Bolyai Research Scholarship, EFOP, the New National Excellence
Program ({\'U}NKP), NKFIH, and OTKA (Hungary),
Department of Atomic Energy and Department of Science and Technology
(India),
Israel Science Foundation (Israel),
Basic Science Research and SRC(CENuM) Programs through NRF
funded by the Ministry of Education and the Ministry of
Science and ICT (Korea).
Physics Department, Lahore University of Management Sciences (Pakistan),
Ministry of Education and Science, Russian Academy of Sciences,
Federal Agency of Atomic Energy (Russia),
VR and Wallenberg Foundation (Sweden),
the U.S. Civilian Research and Development Foundation for the
Independent States of the Former Soviet Union,
the Hungarian American Enterprise Scholarship Fund,
the US-Hungarian Fulbright Foundation,
and the US-Israel Binational Science Foundation.

\end{acknowledgments}

%%%%%%%%%%%%%%%%%%%%%%%%%%%  References 

%\bibliography{ppg238x1}   

\begin{thebibliography}{24}%
\makeatletter
\providecommand \@ifxundefined [1]{%
 \@ifx{#1\undefined}
}%
\providecommand \@ifnum [1]{%
 \ifnum #1\expandafter \@firstoftwo
 \else \expandafter \@secondoftwo
 \fi
}%
\providecommand \@ifx [1]{%
 \ifx #1\expandafter \@firstoftwo
 \else \expandafter \@secondoftwo
 \fi
}%
\providecommand \natexlab [1]{#1}%
\providecommand \enquote  [1]{``#1''}%
\providecommand \bibnamefont  [1]{#1}%
\providecommand \bibfnamefont [1]{#1}%
\providecommand \citenamefont [1]{#1}%
\providecommand \href@noop [0]{\@secondoftwo}%
\providecommand \href [0]{\begingroup \@sanitize@url \@href}%
\providecommand \@href[1]{\@@startlink{#1}\@@href}%
\providecommand \@@href[1]{\endgroup#1\@@endlink}%
\providecommand \@sanitize@url [0]{\catcode `\\12\catcode `\$12\catcode
  `\&12\catcode `\#12\catcode `\^12\catcode `\_12\catcode `\%12\relax}%
\providecommand \@@startlink[1]{}%
\providecommand \@@endlink[0]{}%
\providecommand \url  [0]{\begingroup\@sanitize@url \@url }%
\providecommand \@url [1]{\endgroup\@href {#1}{\urlprefix }}%
\providecommand \urlprefix  [0]{URL }%
\providecommand \Eprint [0]{\href }%
\providecommand \doibase [0]{http://dx.doi.org/}%
\providecommand \selectlanguage [0]{\@gobble}%
\providecommand \bibinfo  [0]{\@secondoftwo}%
\providecommand \bibfield  [0]{\@secondoftwo}%
\providecommand \translation [1]{[#1]}%
\providecommand \BibitemOpen [0]{}%
\providecommand \bibitemStop [0]{}%
\providecommand \bibitemNoStop [0]{.\EOS\space}%
\providecommand \EOS [0]{\spacefactor3000\relax}%
\providecommand \BibitemShut  [1]{\csname bibitem#1\endcsname}%
\let\auto@bib@innerbib\@empty
%</preamble>
\bibitem [{\citenamefont {Adler}\ \emph {et~al.}(2001)\citenamefont {Adler},
  \citenamefont {Denisov}, \citenamefont {Garcia}, \citenamefont {Murray},
  \citenamefont {Strobele},\ and\ \citenamefont {White}}]{Adler:2000bd}%
  \BibitemOpen
  \bibfield  {author} {\bibinfo {author} {\bibfnamefont {C.}~\bibnamefont
  {Adler}}, \bibinfo {author} {\bibfnamefont {A.}~\bibnamefont {Denisov}},
  \bibinfo {author} {\bibfnamefont {E.}~\bibnamefont {Garcia}}, \bibinfo
  {author} {\bibfnamefont {M.~J.}\ \bibnamefont {Murray}}, \bibinfo {author}
  {\bibfnamefont {H.}~\bibnamefont {Strobele}}, \ and\ \bibinfo {author}
  {\bibfnamefont {S.~N.}\ \bibnamefont {White}},\ }\bibfield  {title} {\enquote
  {\bibinfo {title} {{The RHIC zero degree calorimeter}},}\ }\href@noop {}
  {\bibfield  {journal} {\bibinfo  {journal} {Nucl. Instrum. Methods Phys.
  Res., Sec. A}\ }\textbf {\bibinfo {volume} {470}},\ \bibinfo {pages} {488}
  (\bibinfo {year} {2001})}\BibitemShut {NoStop}%
\bibitem [{\citenamefont {Adams}\ \emph {et~al.}(1991)\citenamefont {Adams}
  \emph {et~al.}}]{Adams:1991rw}%
  \BibitemOpen
  \bibfield  {author} {\bibinfo {author} {\bibfnamefont {D.L.}\ \bibnamefont
  {Adams}} \emph {et~al.} (\bibinfo {collaboration} {E581 and E704
  Collaborations}),\ }\bibfield  {title} {\enquote {\bibinfo {title}
  {{Comparison of spin asymmetries and cross-sections in pi0 production by
  200-GeV polarized anti-protons and protons}},}\ }\href {\doibase
  10.1016/0370-2693(91)91351-U} {\bibfield  {journal} {\bibinfo  {journal}
  {Phys. Lett. B}\ }\textbf {\bibinfo {volume} {261}},\ \bibinfo {pages} {201}
  (\bibinfo {year} {1991})}\BibitemShut {NoStop}%
\bibitem [{\citenamefont {Kim}\ \emph {et~al.}(2020)\citenamefont {Kim} \emph
  {et~al.}}]{Kim:2020sxu}%
  \BibitemOpen
  \bibfield  {author} {\bibinfo {author} {\bibfnamefont {M.H.}\ \bibnamefont
  {Kim}} \emph {et~al.} (\bibinfo {collaboration} {RHICf Collaboration}),\
  }\bibfield  {title} {\enquote {\bibinfo {title} {{Transverse single-spin
  asymmetry for very forward neutral pion production in polarized $p+p$
  collisions at $\sqrt{s} = 510$ GeV}},}\ }\href@noop {} {\bibfield  {journal}
  {\bibinfo  {journal} {Phys. Rev. Lett.}\ }\textbf {\bibinfo {volume} {124}},\
  \bibinfo {pages} {252501} (\bibinfo {year} {2020})}\BibitemShut {NoStop}%
\bibitem [{\citenamefont {Fukao}\ \emph {et~al.}(2007)\citenamefont {Fukao}
  \emph {et~al.}}]{Bazilevsky:2006vd}%
  \BibitemOpen
  \bibfield  {author} {\bibinfo {author} {\bibfnamefont {Y.}~\bibnamefont
  {Fukao}} \emph {et~al.},\ }\bibfield  {title} {\enquote {\bibinfo {title}
  {{Single Transverse-Spin Asymmetry in Very Forward and Very Backward Neutral
  Particle Production for Polarized Proton Collisions at s**(1/2) =
  200-GeV}},}\ }\href@noop {} {\bibfield  {journal} {\bibinfo  {journal} {Phys.
  Lett. B}\ }\textbf {\bibinfo {volume} {650}},\ \bibinfo {pages} {325}
  (\bibinfo {year} {2007})}\BibitemShut {NoStop}%
\bibitem [{\citenamefont {Soffer}\ and\ \citenamefont
  {Tornqvist}(1992)}]{Soffer:1991am}%
  \BibitemOpen
  \bibfield  {author} {\bibinfo {author} {\bibfnamefont {J.}~\bibnamefont
  {Soffer}}\ and\ \bibinfo {author} {\bibfnamefont {N.~A.}\ \bibnamefont
  {Tornqvist}},\ }\bibfield  {title} {\enquote {\bibinfo {title} {{Origin of
  the polarization for inclusive lambda production in p p collisions}},}\
  }\href {\doibase 10.1103/PhysRevLett.68.907} {\bibfield  {journal} {\bibinfo
  {journal} {Phys. Rev. Lett.}\ }\textbf {\bibinfo {volume} {68}},\ \bibinfo
  {pages} {907} (\bibinfo {year} {1992})}\BibitemShut {NoStop}%
\bibitem [{\citenamefont {D'Alesio}\ and\ \citenamefont
  {Pirner}(2000)}]{DAlesio:1998uav}%
  \BibitemOpen
  \bibfield  {author} {\bibinfo {author} {\bibfnamefont {U.}~\bibnamefont
  {D'Alesio}}\ and\ \bibinfo {author} {\bibfnamefont {H.~J.}\ \bibnamefont
  {Pirner}},\ }\bibfield  {title} {\enquote {\bibinfo {title} {{Target
  fragmentation in p p, e p and gamma p collisions at high-energies}},}\
  }\href@noop {} {\bibfield  {journal} {\bibinfo  {journal} {Eur. Phys. J. A}\
  }\textbf {\bibinfo {volume} {7}},\ \bibinfo {pages} {109} (\bibinfo {year}
  {2000})}\BibitemShut {NoStop}%
\bibitem [{\citenamefont {Kopeliovich}\ \emph {et~al.}(1996)\citenamefont
  {Kopeliovich}, \citenamefont {Povh},\ and\ \citenamefont
  {Potashnikova}}]{Kopeliovich:1996iw}%
  \BibitemOpen
  \bibfield  {author} {\bibinfo {author} {\bibfnamefont {B.}~\bibnamefont
  {Kopeliovich}}, \bibinfo {author} {\bibfnamefont {B.}~\bibnamefont {Povh}}, \
  and\ \bibinfo {author} {\bibfnamefont {I.}~\bibnamefont {Potashnikova}},\
  }\bibfield  {title} {\enquote {\bibinfo {title} {{Deep inelastic
  electroproduction of neutrons in the proton fragmentation region}},}\
  }\href@noop {} {\bibfield  {journal} {\bibinfo  {journal} {Z. Phys. C}\
  }\textbf {\bibinfo {volume} {73}},\ \bibinfo {pages} {125} (\bibinfo {year}
  {1996})}\BibitemShut {NoStop}%
\bibitem [{\citenamefont {Flauger}\ and\ \citenamefont
  {Monnig}(1976)}]{Flauger:1976ju}%
  \BibitemOpen
  \bibfield  {author} {\bibinfo {author} {\bibfnamefont {W.}~\bibnamefont
  {Flauger}}\ and\ \bibinfo {author} {\bibfnamefont {F.}~\bibnamefont
  {Monnig}},\ }\bibfield  {title} {\enquote {\bibinfo {title} {{Measurement of
  Inclusive Zero-Angle Neutron Spectra at the ISR}},}\ }\href@noop {}
  {\bibfield  {journal} {\bibinfo  {journal} {Nucl. Phys.}\ }\textbf {\bibinfo
  {volume} {B109}},\ \bibinfo {pages} {347} (\bibinfo {year}
  {1976})}\BibitemShut {NoStop}%
\bibitem [{\citenamefont {Collins}(2009)}]{Collins:1977jy}%
  \BibitemOpen
  \bibfield  {author} {\bibinfo {author} {\bibfnamefont {P.D.B.}\ \bibnamefont
  {Collins}},\ }\href {\doibase 10.1017/CBO9780511897603} {\emph {\bibinfo
  {title} {{An Introduction to Regge Theory and High-Energy Physics}}}},\
  Cambridge Monographs on Mathematical Physics\ (\bibinfo  {publisher}
  {Cambridge Univ. Press},\ \bibinfo {address} {Cambridge, UK},\ \bibinfo
  {year} {2009})\BibitemShut {NoStop}%
\bibitem [{\citenamefont {Kopeliovich}\ \emph {et~al.}(2011)\citenamefont
  {Kopeliovich}, \citenamefont {Potashnikova}, \citenamefont {Schmidt},\ and\
  \citenamefont {Soffer}}]{Kopeliovich:2011bx}%
  \BibitemOpen
  \bibfield  {author} {\bibinfo {author} {\bibfnamefont {B.Z.}\ \bibnamefont
  {Kopeliovich}}, \bibinfo {author} {\bibfnamefont {I.K.}\ \bibnamefont
  {Potashnikova}}, \bibinfo {author} {\bibfnamefont {Ivan}\ \bibnamefont
  {Schmidt}}, \ and\ \bibinfo {author} {\bibfnamefont {J.}~\bibnamefont
  {Soffer}},\ }\bibfield  {title} {\enquote {\bibinfo {title} {{Single
  transverse spin asymmetry of forward neutrons}},}\ }\href {\doibase
  10.1103/PhysRevD.84.114012} {\bibfield  {journal} {\bibinfo  {journal} {Phys.
  Rev. D}\ }\textbf {\bibinfo {volume} {84}},\ \bibinfo {pages} {114012}
  (\bibinfo {year} {2011})}\BibitemShut {NoStop}%
\bibitem [{\citenamefont {Adare}\ \emph {et~al.}(2013)\citenamefont {Adare}
  \emph {et~al.}}]{Adare:2012vw}%
  \BibitemOpen
  \bibfield  {author} {\bibinfo {author} {\bibfnamefont {A.}~\bibnamefont
  {Adare}} \emph {et~al.} (\bibinfo {collaboration} {PHENIX Collaboration}),\
  }\bibfield  {title} {\enquote {\bibinfo {title} {{Inclusive cross section and
  single transverse spin asymmetry for very forward neutron production in
  polarized p+p collisions at s=200 GeV}},}\ }\href@noop {} {\bibfield
  {journal} {\bibinfo  {journal} {Phys. Rev. D}\ }\textbf {\bibinfo {volume}
  {88}},\ \bibinfo {pages} {032006} (\bibinfo {year} {2013})}\BibitemShut
  {NoStop}%
\bibitem [{\citenamefont {Aidala}\ \emph {et~al.}(2018)\citenamefont {Aidala}
  \emph {et~al.}}]{Aidala:2017cnz}%
  \BibitemOpen
  \bibfield  {author} {\bibinfo {author} {\bibfnamefont {C.}~\bibnamefont
  {Aidala}} \emph {et~al.} (\bibinfo {collaboration} {PHENIX Collaboration}),\
  }\bibfield  {title} {\enquote {\bibinfo {title} {{Nuclear Dependence of the
  Transverse-Single-Spin Asymmetry for Forward Neutron Production in Polarized
  $p+A$ Collisions at $\sqrt{{s}_{NN}}=200$ GeV}},}\ }\href@noop {} {\bibfield
  {journal} {\bibinfo  {journal} {Phys. Rev. Lett.}\ }\textbf {\bibinfo
  {volume} {120}},\ \bibinfo {pages} {022001} (\bibinfo {year}
  {2018})}\BibitemShut {NoStop}%
\bibitem [{\citenamefont {Mitsuka}(2017)}]{Mitsuka:2017czj}%
  \BibitemOpen
  \bibfield  {author} {\bibinfo {author} {\bibfnamefont {G.}~\bibnamefont
  {Mitsuka}},\ }\bibfield  {title} {\enquote {\bibinfo {title} {{Recently
  measured large $A_N$ for forward neutrons in $p^\uparrow$$+$$A$ collisions at
  $\sqrt{s_{NN}}=200$ GeV explained through simulations of ultraperipheral
  collisions and hadronic interactions}},}\ }\href@noop {} {\bibfield
  {journal} {\bibinfo  {journal} {Phys. Rev. C}\ }\textbf {\bibinfo {volume}
  {95}},\ \bibinfo {pages} {044908} (\bibinfo {year} {2017})}\BibitemShut
  {NoStop}%
\bibitem [{\citenamefont {Adcox}\ \emph {et~al.}(2003)\citenamefont {Adcox}
  \emph {et~al.}}]{Adcox:2003zm}%
  \BibitemOpen
  \bibfield  {author} {\bibinfo {author} {\bibfnamefont {K.}~\bibnamefont
  {Adcox}} \emph {et~al.} (\bibinfo {collaboration} {PHENIX Collaboration}),\
  }\bibfield  {title} {\enquote {\bibinfo {title} {{PHENIX detector
  overview}},}\ }\href {\doibase 10.1016/S0168-9002(02)01950-2} {\bibfield
  {journal} {\bibinfo  {journal} {Nucl. Instrum. Methods Phys. Res., Sec. A}\
  }\textbf {\bibinfo {volume} {499}},\ \bibinfo {pages} {469} (\bibinfo {year}
  {2003})}\BibitemShut {NoStop}%
\bibitem [{\citenamefont {{RHIC Polarimetry Group}}(2013)}]{polarimetry}%
  \BibitemOpen
  \bibfield  {author} {\bibinfo {author} {\bibnamefont {{RHIC Polarimetry
  Group}}},\ }\href@noop {} {\enquote {\bibinfo {title} {{RHIC polarization for
  Runs 9--12, RHIC/CAD Accelerator Physics Note 490}},}\ } (\bibinfo {year}
  {2013})\BibitemShut {NoStop}%
\bibitem [{\citenamefont {Brun}\ \emph {et~al.}(1994)\citenamefont {Brun},
  \citenamefont {Bruyant}, \citenamefont {Carminati}, \citenamefont {Giani},
  \citenamefont {Maire}, \citenamefont {McPherson}, \citenamefont {Patrick},\
  and\ \citenamefont {Urban}}]{Brun:1994aa}%
  \BibitemOpen
  \bibfield  {author} {\bibinfo {author} {\bibfnamefont {R.}~\bibnamefont
  {Brun}}, \bibinfo {author} {\bibfnamefont {F.}~\bibnamefont {Bruyant}},
  \bibinfo {author} {\bibfnamefont {F.}~\bibnamefont {Carminati}}, \bibinfo
  {author} {\bibfnamefont {S.}~\bibnamefont {Giani}}, \bibinfo {author}
  {\bibfnamefont {M.}~\bibnamefont {Maire}}, \bibinfo {author} {\bibfnamefont
  {A.}~\bibnamefont {McPherson}}, \bibinfo {author} {\bibfnamefont
  {G.}~\bibnamefont {Patrick}}, \ and\ \bibinfo {author} {\bibfnamefont
  {L.}~\bibnamefont {Urban}},\ }\href@noop {} {\enquote {\bibinfo {title}
  {{GEANT Detector Description and Simulation Tool, CERN-W5013, CERN-W-5013,
  W5013, W-5013}},}\ } (\bibinfo {year} {1994})\BibitemShut {NoStop}%
\bibitem [{\citenamefont {Togawa}(2008)}]{Togawa:2008cca}%
  \BibitemOpen
  \bibfield  {author} {\bibinfo {author} {\bibfnamefont {M.}~\bibnamefont
  {Togawa}},\ }\emph {\bibinfo {title} {{Measurements of the leading neutron
  production in polarized $pp$ collision at $\sqrt{s}$=200 GeV}}},\ \href@noop
  {} {Ph.D. thesis},\ \bibinfo  {school} {Kyoto University} (\bibinfo {year}
  {2008})\BibitemShut {NoStop}%
\bibitem [{\citenamefont {Adare}\ \emph {et~al.}(2014)\citenamefont {Adare}
  \emph {et~al.}}]{Adare:2013ekj}%
  \BibitemOpen
  \bibfield  {author} {\bibinfo {author} {\bibfnamefont {A.}~\bibnamefont
  {Adare}} \emph {et~al.} (\bibinfo {collaboration} {PHENIX Collaboration}),\
  }\bibfield  {title} {\enquote {\bibinfo {title} {{Measurement of
  transverse-single-spin asymmetries for midrapidity and forward-rapidity
  production of hadrons in polarized p+p collisions at $\sqrt{s}=$200 and 62.4
  GeV}},}\ }\href@noop {} {\bibfield  {journal} {\bibinfo  {journal} {Phys.
  Rev. D}\ }\textbf {\bibinfo {volume} {90}},\ \bibinfo {pages} {012006}
  (\bibinfo {year} {2014})}\BibitemShut {NoStop}%
\bibitem [{\citenamefont {Sjostrand}\ \emph {et~al.}(2001)\citenamefont
  {Sjostrand}, \citenamefont {Lonnblad},\ and\ \citenamefont
  {Mrenna}}]{Sjostrand:2001yu}%
  \BibitemOpen
  \bibfield  {author} {\bibinfo {author} {\bibfnamefont {T.}~\bibnamefont
  {Sjostrand}}, \bibinfo {author} {\bibfnamefont {L.}~\bibnamefont {Lonnblad}},
  \ and\ \bibinfo {author} {\bibfnamefont {S.}~\bibnamefont {Mrenna}},\
  }\href@noop {} {\enquote {\bibinfo {title} {{PYTHIA 6.2: Physics and
  manual}},}\ } (\bibinfo {year} {2001})\BibitemShut {NoStop}%
\bibitem [{\citenamefont {Sj\"ostrand}\ \emph {et~al.}(2015)\citenamefont
  {Sj\"ostrand}, \citenamefont {Ask}, \citenamefont {Christiansen},
  \citenamefont {Corke}, \citenamefont {Desai}, \citenamefont {Ilten},
  \citenamefont {Mrenna}, \citenamefont {Prestel}, \citenamefont {Rasmussen},\
  and\ \citenamefont {Skands}}]{Sjostrand:2014zea}%
  \BibitemOpen
  \bibfield  {author} {\bibinfo {author} {\bibfnamefont {T.}~\bibnamefont
  {Sj\"ostrand}}, \bibinfo {author} {\bibfnamefont {S.}~\bibnamefont {Ask}},
  \bibinfo {author} {\bibfnamefont {J.~R.}\ \bibnamefont {Christiansen}},
  \bibinfo {author} {\bibfnamefont {R.}~\bibnamefont {Corke}}, \bibinfo
  {author} {\bibfnamefont {N.}~\bibnamefont {Desai}}, \bibinfo {author}
  {\bibfnamefont {P.}~\bibnamefont {Ilten}}, \bibinfo {author} {\bibfnamefont
  {S.}~\bibnamefont {Mrenna}}, \bibinfo {author} {\bibfnamefont
  {S.}~\bibnamefont {Prestel}}, \bibinfo {author} {\bibfnamefont
  {Christine~O.}\ \bibnamefont {Rasmussen}}, \ and\ \bibinfo {author}
  {\bibfnamefont {Peter~Z.}\ \bibnamefont {Skands}},\ }\bibfield  {title}
  {\enquote {\bibinfo {title} {{An introduction to PYTHIA 8.2}},}\ }\href
  {\doibase 10.1016/j.cpc.2015.01.024} {\bibfield  {journal} {\bibinfo
  {journal} {Comput. Phys. Commun.}\ }\textbf {\bibinfo {volume} {191}},\
  \bibinfo {pages} {159} (\bibinfo {year} {2015})}\BibitemShut {NoStop}%
\bibitem [{\citenamefont {Roesler}\ \emph {et~al.}(2000)\citenamefont
  {Roesler}, \citenamefont {Engel},\ and\ \citenamefont
  {Ranft}}]{Roesler:2000he}%
  \BibitemOpen
  \bibfield  {author} {\bibinfo {author} {\bibfnamefont {S.}~\bibnamefont
  {Roesler}}, \bibinfo {author} {\bibfnamefont {R.}~\bibnamefont {Engel}}, \
  and\ \bibinfo {author} {\bibfnamefont {J.}~\bibnamefont {Ranft}},\ }\bibfield
   {title} {\enquote {\bibinfo {title} {{The Monte Carlo event generator
  DPMJET-III}},}\ }in\ \href {\doibase 10.1007/978-3-642-18211-2_166} {\emph
  {\bibinfo {booktitle} {{International Conference on Advanced Monte Carlo for
  Radiation Physics, Particle Transport Simulation and Applications (MC
  2000)}}}}\ (\bibinfo {year} {2000})\ p.\ \bibinfo {pages} {1033},\ \Eprint
  {http://arxiv.org/abs/hep-ph/0012252} {arXiv:hep-ph/0012252} \BibitemShut
  {NoStop}%
\bibitem [{\citenamefont {Klein}\ \emph {et~al.}(2017)\citenamefont {Klein},
  \citenamefont {Nystrand}, \citenamefont {Seger}, \citenamefont {Gorbunov},\
  and\ \citenamefont {Butterworth}}]{Klein:2016yzr}%
  \BibitemOpen
  \bibfield  {author} {\bibinfo {author} {\bibfnamefont {S.~R.}\ \bibnamefont
  {Klein}}, \bibinfo {author} {\bibfnamefont {J.}~\bibnamefont {Nystrand}},
  \bibinfo {author} {\bibfnamefont {J.}~\bibnamefont {Seger}}, \bibinfo
  {author} {\bibfnamefont {Y.}~\bibnamefont {Gorbunov}}, \ and\ \bibinfo
  {author} {\bibfnamefont {J.}~\bibnamefont {Butterworth}},\ }\bibfield
  {title} {\enquote {\bibinfo {title} {{STARlight: A Monte Carlo simulation
  program for ultra-peripheral collisions of relativistic ions}},}\ }\href@noop
  {} {\bibfield  {journal} {\bibinfo  {journal} {Comput. Phys. Commun.}\
  }\textbf {\bibinfo {volume} {212}},\ \bibinfo {pages} {258} (\bibinfo {year}
  {2017})}\BibitemShut {NoStop}%
\bibitem [{\citenamefont {Brun}\ and\ \citenamefont
  {Rademakers}(1997)}]{Brun:1997pa}%
  \BibitemOpen
  \bibfield  {author} {\bibinfo {author} {\bibfnamefont {R.}~\bibnamefont
  {Brun}}\ and\ \bibinfo {author} {\bibfnamefont {F.}~\bibnamefont
  {Rademakers}},\ }\bibfield  {title} {\enquote {\bibinfo {title} {{ROOT: An
  object oriented data analysis framework}},}\ }\href {\doibase
  10.1016/S0168-9002(97)00048-X} {\bibfield  {journal} {\bibinfo  {journal}
  {Nucl. Instrum. Methods Phys. Res., Sec. A}\ }\textbf {\bibinfo {volume}
  {389}},\ \bibinfo {pages} {81} (\bibinfo {year} {1997})}\BibitemShut
  {NoStop}%
\bibitem [{\citenamefont {Hocker}\ and\ \citenamefont
  {Kartvelishvili}(1996)}]{Hocker:1995kb}%
  \BibitemOpen
  \bibfield  {author} {\bibinfo {author} {\bibfnamefont {A.}~\bibnamefont
  {Hocker}}\ and\ \bibinfo {author} {\bibfnamefont {V.}~\bibnamefont
  {Kartvelishvili}},\ }\bibfield  {title} {\enquote {\bibinfo {title} {{SVD
  approach to data unfolding}},}\ }\href@noop {} {\bibfield  {journal}
  {\bibinfo  {journal} {Nucl. Instrum. Methods Phys. Res., Sec. A}\ }\textbf
  {\bibinfo {volume} {372}},\ \bibinfo {pages} {469} (\bibinfo {year}
  {1996})}\BibitemShut {NoStop}%
\end{thebibliography}

 
\end{document}